# Wavelength-Resolved Control of Photovoltaic Screening and Defect-Mediated Doping in Photo-Ferroelectric/Graphene Devices

Krishna Prasad Maity,* Mohd Uvais, Jean-François Dayen, Bernard Doudin, Roman Gumeniuk, and Bohdan Kundys*

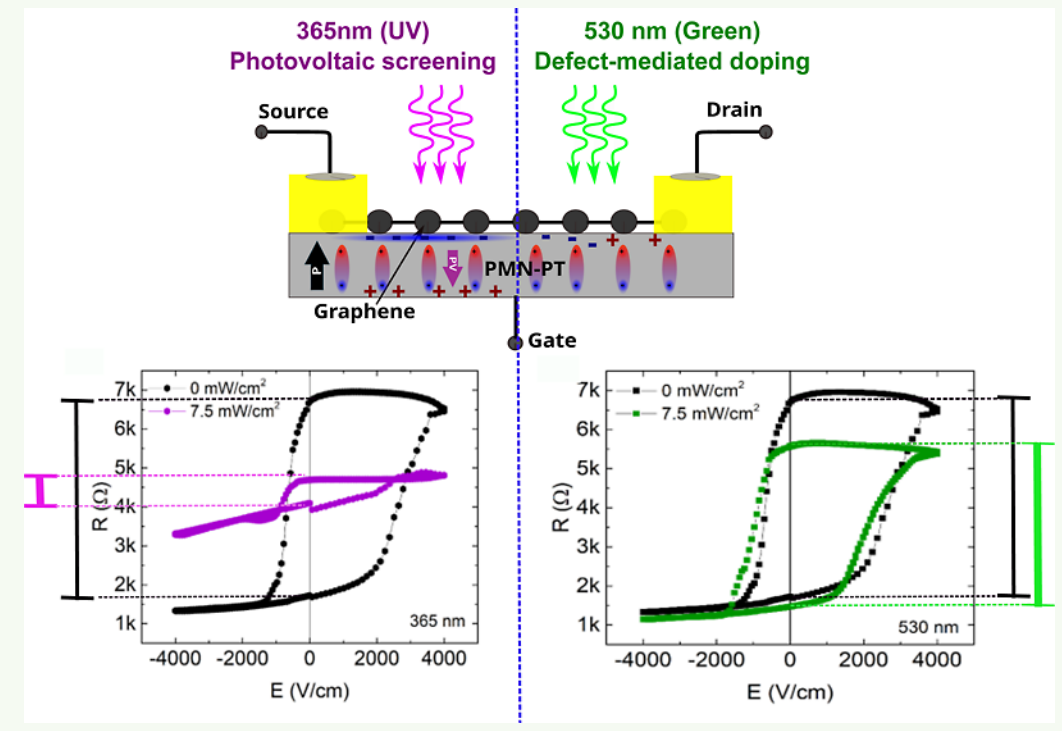

**ABSTRACT:** Ferroelectrics enable large charge doping of two-dimensional overlayers, but the coexistence of switching and nonswitching charge dynamics complicate electro-optical analysis. Here, we investigate the optoelectronic response of a ferroelectric/graphene device under 365 and 530 nm illumination, disentangling effects on ferroelectric dipole alignment from extrinsic current pathways. Graphene acts as a high-gain sensor, amplifying subtle polarization dynamics into a pronounced resistance difference. By resolving switching and nonswitching channels in dark and illuminated conditions, we reveal a competition between photovoltaic charge screening and defect-assisted excitation that governs device electrostatics. Above-band gap illumination generates free carriers that induce leaky ferroelectric hysteresis and suppress the graphene resistance ratio between opposite remanent polarization states from 290% to 15% due to dynamic photovoltaic charge screening. In contrast, 530 nm illumination primarily induces charge redistribution in ferroelectrics via defect-state excitation, leading to a significantly weaker suppression of the resistance variation of graphene. These results establish practical guidelines for selecting photon energy and intensity to either preserve remanent polarization while tuning channel doping or deliberately reconfigure polarization through optical programming.

## 1. INTRODUCTION

Ferroelectric (FE) materials are promising candidates for applications as low-energy memory devices, detectors, and gate dielectrics in transistors.[1−3] Their ability to retain charges at remanent polarization states without an external gate voltage makes them highly suitable for future electronics. The flexibility of tuning FE polarization via electrical and optical excitation enables their use in optoelectronic devices with low-power consumption and fast response times.[4] The diverse range of materials from bulk to 2D or sliding 2D layers exhibiting FE behavior make them attractive for low energy computation applications.[3,5]

Optical modulation of FE polarization introduces an additional degree of freedom for integrating them into FE field effect transistors with 2D semiconductors.[6] Above band gap illumination can alter FE polarization, change internal electric fields, and affect devices built with 2D semiconductors.[7−10] Photogenerated charge carriers within the FE material modulate the screening charge density and polarization, thereby influencing the channel transport in devices. Such optical modulation of ferroelectricity has potential applications in neuromorphic computing, optically writable memory devices, and photodetectors.[11−14] However, the impact of light excitation on channel conductivity and ferroelectric polarization remains largely unexplored in the literature. By investigating the model FE/graphene structure with a clean interface, this study aims to provide insights into many possible light-induced interactions in FE/2D heterostructures and details how channel conductivity can be modulated in FE field-effect transistors under optical excitations. We also detail how it is key to distinguish between "switching", related to FE hysteretic polarization, and "non-switching", related to leakage currents and stray capacitance, when measuring the current in a device. These two contributions influence the measured current in the 2D overlayer in strikingly different ways. This can then better explain photocurrent responses and the charge dynamics behavior.

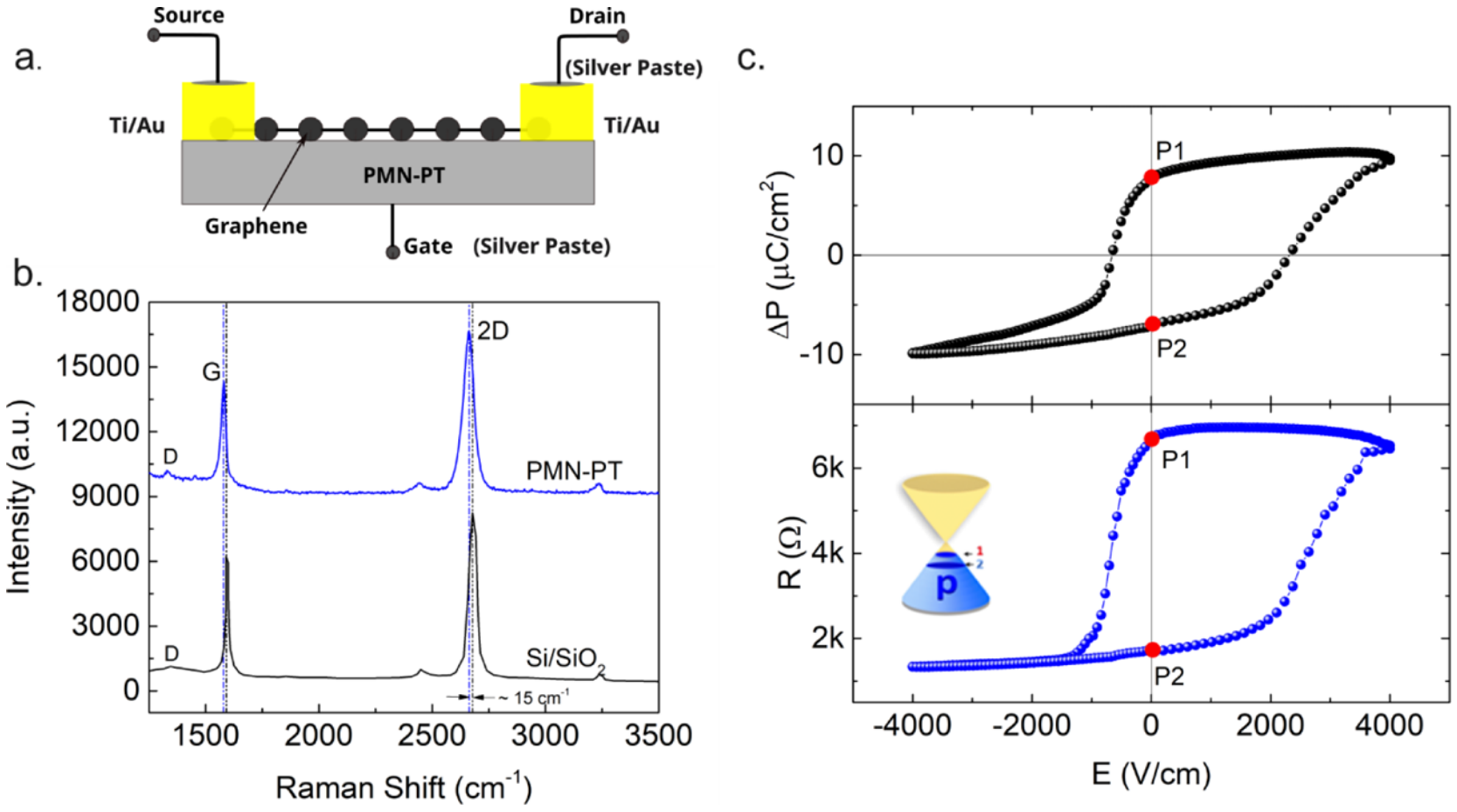

**Figure 1.** (a) Schematic of the graphene/ferroelectric device, (b) Raman spectrum for monolayer graphene on top of $Si/SiO_2$ and PMN-PT substrates, and (c) FE loop and graphene resistance variation with applied gate electric field.

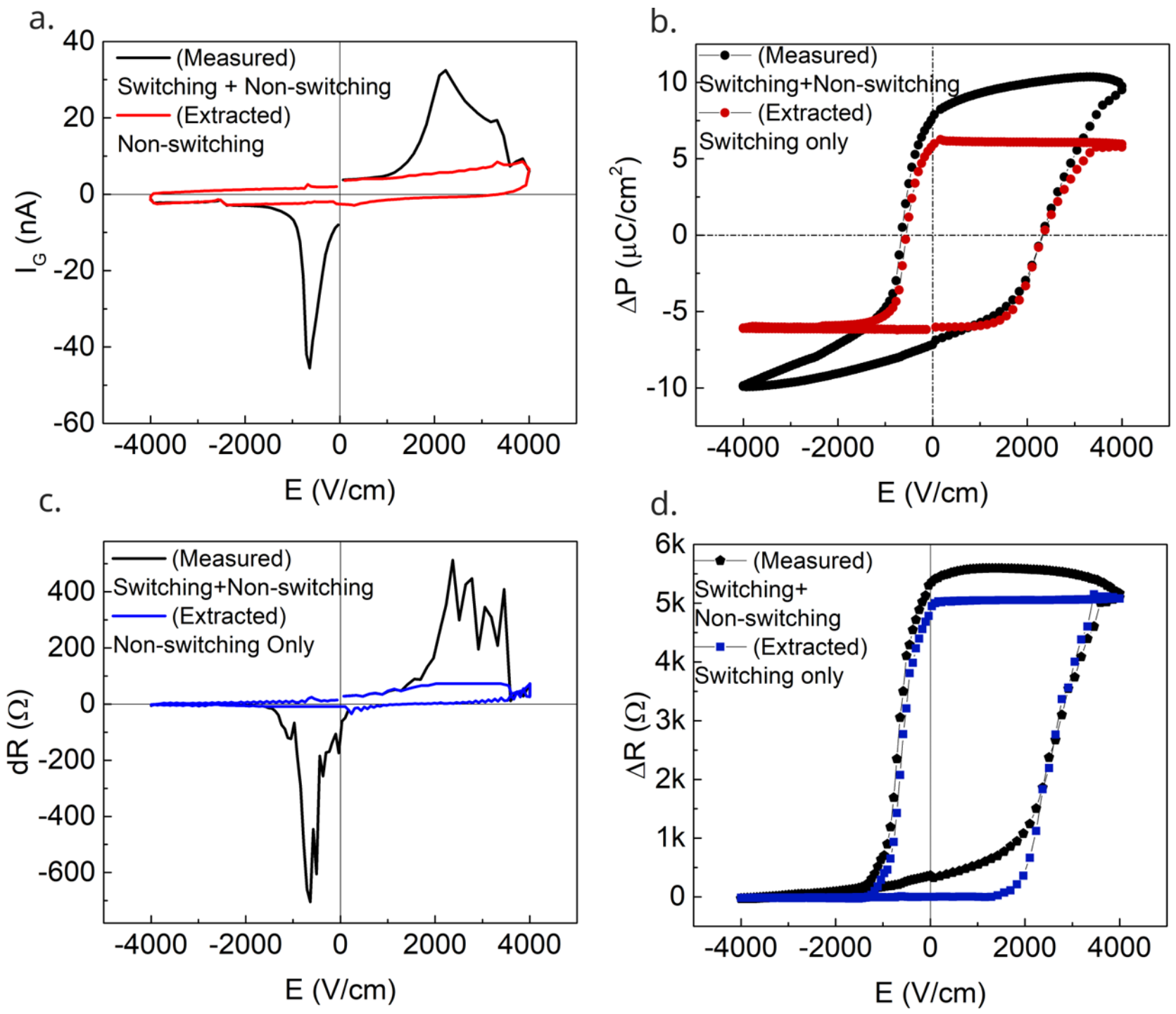

**Figure 2.** Measured and corrected components of source gate current (a) related ferroelectric loop (b) which disentangle the nonswitching contribution. (c,d) Same analysis applied to distinguish switching and nonswitching effects on graphene transconductance measured in situ (see the text).

## 2. EXPERIMENTAL SECTION

The $Pb[(Mg_{1/3}Nb_{2/3})_{0.70}Ti_{0.30}]O_3$ crystal with dimensions of $0.5 \times 0.05 \times 0.03$ cm$^3$ was cut from plates of (001) orientation supplied by Crystal GmbH (Germany), and monolayer of graphene was grown on top of the crystal using chemical vapor deposition by Graphenea. The two top electrodes were made by depositing Ti (10 nm)/Au (50 nm) on graphene, and silver paste was used for the bottom electrode, as shown in Figure 1a. Quasi-static ferroelectric measurements were performed at 5 mHz to reduce ferroelectric fatigue risks.

## 3. RESULTS AND DISCUSSION

### 3.1. Ground-State Doping and FE Doping of Graphene

The Raman spectrum of graphene on $Si/SiO_2$ and PMN-PT substrates is shown in Figure 1b. Dominant peaks at 1591 cm$^{-1}$ and 2678 cm$^{-1}$ correspond to the G peak and 2D peak of monolayer graphene, respectively. The peaks are blue-shifted by ∼15 cm$^{-1}$, indicating the strong interaction and p-type of charge doping in graphene due to PMN-PT substrate, in agreement with previous work.[15] Compared to graphene on $SiO_2$ from the same bunch, graphene on the ferroelectric substrate exhibits correlated softening of the G and 2D Raman modes with an effective slope $\Delta\omega_{2D}/\Delta\omega_G \approx 1.5$, indicating a mixed response dominated by ferroelectric-induced charge doping with a secondary strain contribution.[16]

In Figure 1c, the FE hysteresis loop with applied gate voltage is mimicked by the graphene resistance hysteresis.[17,18] The variation in resistance with gate voltage closely follows the ferroelectric loop behavior, exhibiting a significant resistance difference: $\Delta R = [(R(P1)\text{-}R(P2))/R(P1)]\times100\%$; $R(P1)$ and $R(P2)$ represent the resistance at positive and negative remanent polarization states ($E = 0$ V/cm), respectively. In the dark, $\Delta R$ reaches ~290%. It has to be noted that the CVD graphene deposition procedure has been found to reduce the remanent polarization of the PMN-PT substrate by more than three times, most likely due to thermal history of the FE sample.[19] Graphene's Fermi level is tuned by the FE polarization via gate voltage, and it shows a higher resistance value closer to the Dirac point (shown in the inset). A clear transconductance hysteresis is dominated by the switchable FE polarization, though both switching and nonswitching currents are present, testifying to the p-type strong doping regime far below the Dirac point. For positive ferroelectric poling (P1), the graphene resistance increases to approximately 7.0 kΩ due to electron doping, whereas negative poling (P2) induces a larger population of hole carriers, resulting in much lower resistance values (~1.5 kΩ).

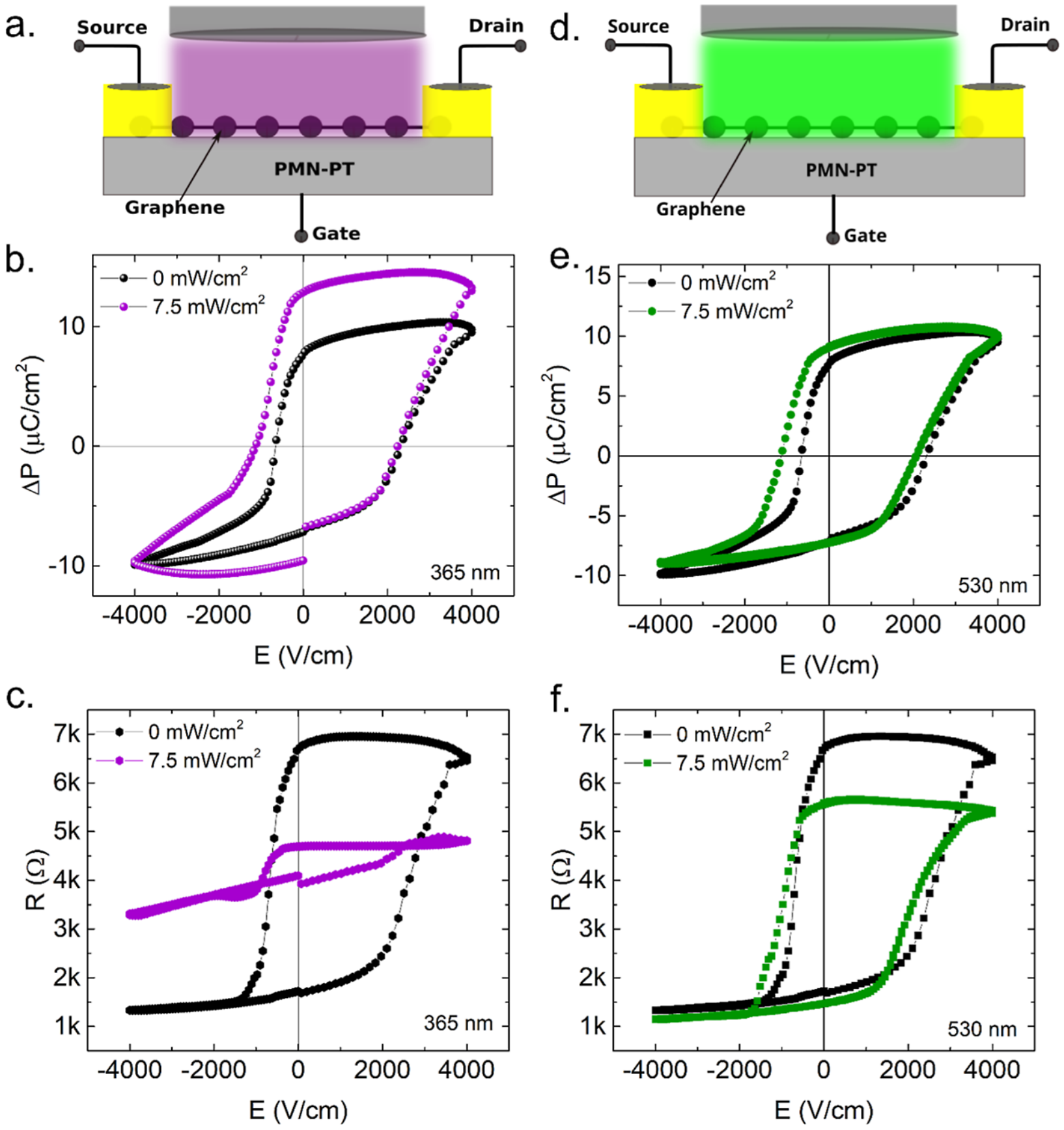

**Figure 3.** (a), (d) Schematic of the PMN-PT/graphene device with optical irradiation of UV and green light, (b) and (c) polarization and graphene resistance variations with gate voltage in the presence of continuous UV light (365 nm). (e,f) polarization and graphene transconductance in the presence of continuous green light (530 nm).

## 3.2. Disentangling Switching and Nonswitching Contributions

**3.2.1. Ferroelectric and Graphene FE Doping Analysis.** Due to undesired imperfection of dielectric materials resulting in an electric leakage, the non-zero DC conductivity may hinder the FE switching under electric field.[20] This ohmic semiconductive behavior increases the total current value as a function of applied electric field and therefore also impacts the graphene overlayer properties. Nonswitching currents can result from both: paraelectric atomic displacements (at large electric fields) and space charges currents, and distinguishing them from pure dipole FE switching is an important task.[21] Therefore, an insightful analysis of the FE ground state should be provided in the dark to distinguish electric leakage doping contribution from FE doping, when analyzing the electric field-induced effect in 2D/FE heterostructures. The equation describing current during electric field cycling can be written as follows

$$i_t = A\left(\varepsilon_0 \frac{\partial E_{ex}}{\partial t} + \frac{\partial P_s}{\partial t} + \frac{\partial P_{ns}}{\partial t}\right) \quad (1)$$

where $A$ is a graphene electrode area, $E_{ex}$ is the applied external electric field, and $P_s$ and $P_{ns}$ are switching and nonswitching polarizations, respectively. To distinguish these contributions, a so-called positive-up negative-down (PUND) or "Double Wave" methods can be applied for comparing switching and nonswitching FE. Figure 2a illustrates how the nonswitching background (red color) contributes to the total current measurements (black) measured in the dark. Full explanations and data are provided in the Supporting Information.

The integration of the PUND correction results in a more saturated ferroelectric loop, as shown in Figure 2b (details in Supporting Information). Even a small nonswitching contribution may lead to large overestimation of both remanent FE polarization and coercive field. Moreover, when working with 2D structures, the sample's capacitance can be comparable to or even smaller than the stray capacitance of the electrical circuit, and this must be taken into account when modeling the system with ideal circuit elements. The eq 1

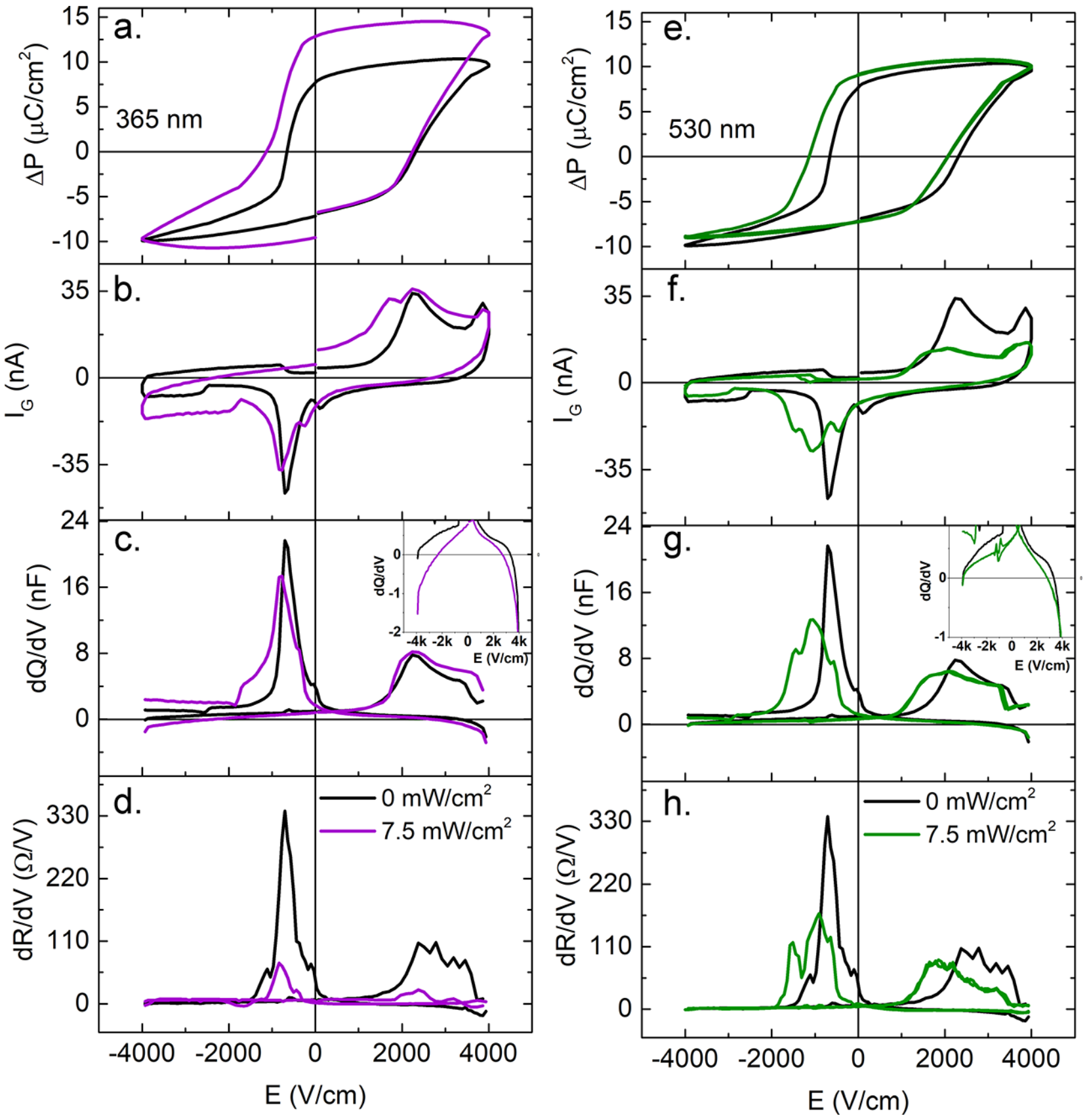

**Figure 4.** Variation of FE polarization, gate FE current, capacitance, and graphene resistance changes with electric field for (a–d) 365 nm UV and (e–h) 530 nm green light irradiations.

should then be modified to consider these contributions, as follows

$$i_t = A\left(\varepsilon_0 \frac{\partial E_{ex}}{\partial t} + \frac{\partial P_s}{\partial t} + \frac{\partial P_{ns}}{\partial t}\right) + C_s \frac{\partial V}{\partial t} \tag{2}$$

or changing $E_{ex} = -V/d$, where $d$ is the thickness of the FE and $V$ is an applied voltage. We get

$$i_t = \left(-\frac{A\epsilon_0}{d} + C_s\right)\frac{\partial V}{\partial t} + A\left(\frac{\partial P_n}{\partial t} + \frac{\partial P_{ns}}{\partial t}\right) \tag{3}$$

The nonzero extra capacitance $C_s$ in the circuit can also have detrimental effects on the FE loop of 2D-based FEs, making a similar artificial opening of the gap in I–V curve near zero voltage (see Figure 2a) and distinguishing its contribution can be a difficult task. However, 2D graphene transport is not affected by this effect as it is only sensitive to surface charges. This key-property provides better insight into the FE analysis confirming, or not, an intrinsic nature of nonswitching currents in FE/2D heterostructures.

It can therefore be proposed that dynamic graphene doping during the FE cycle can be analyzed in the same manner as in the case of I–V analysis (Figure 2c). The differentiated resistance plot as a function of voltage exhibits nonzero values similar to those observed in the nonswitching loop (blue curve in Figure 2c), confirming the intrinsic nature of graphene doping by nonswitching currents in our sample. In agreement with FE loop analysis, the nonswitching part is larger for positive voltages (Figure 2d). After subtracting the nonswitching part of differentiated graphene resistance from the "as-recorded" data, its integration reveals the "FE switching only effect" on the graphene transconductance (Figure 2d). As in the case of polarization loops, presence of nonswitching currents enlarges the remanent resistance values and increases the negative coercive voltage due to asymmetric doping.

## 4. LIGHT EFFECT ON THE FE LOOP AND GRAPHENE TRANSCONDUCTANCE

Following analysis of the effects of both switching and nonswitching FE contributions on graphene transconductance under dark conditions, a similar approach can be applied to the characterization of FE hysteresis loops under illumination. Depending on the photon energy relative to the FE band gap, both the switching and nonswitching components of the ferroelectric response may be modified. Consequently, the graphene transconductance may also be affected, as discussed in the following sections. In Figure 3b, the polarization of PMN-PT exhibits a FE loop with coercive fields ($E_c^+$ and $E_c^-$) of 2372 and −638 V/cm in the absence of light. The shift of the FE loop toward positive voltage suggests the presence of a strong built-in field [$E_{int} = (E_c^+ + E_c^-)/2$] within the FE, with insufficient charge screening by graphene.[22−25] As the UV light irradiates, the coercive fields decrease to 2237 and −1109 V/cm, resulting in a more symmetric loop (Figure 3b). Since UV light with a wavelength of 365 nm and energy $E_{365}$ = 3.4 eV—higher than the PMN-PT band gap ($E_g$ = 3.1 eV)—it induces free charge generation through band-to-band electronic transitions. These photogenerated charges enhance polarization screening, reducing the built-in electric field and leading to a more symmetric FE loop (see Table S1 in Supporting Information). Additionally, the apparent remanent polarization gradually increases due to the accumulation of free charges and enhanced current within the FE material.[23,26,27]

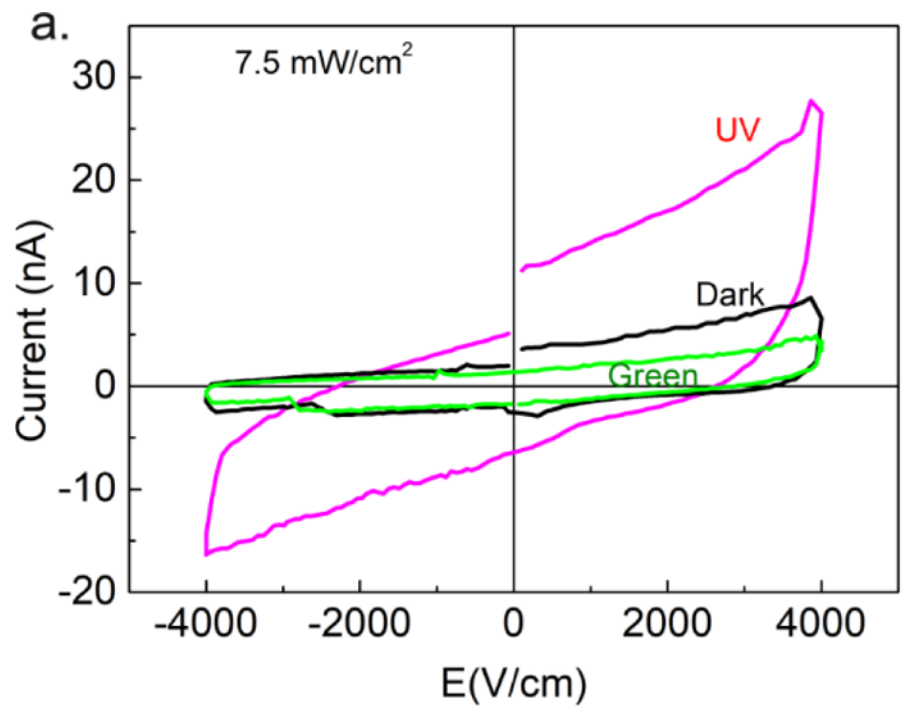

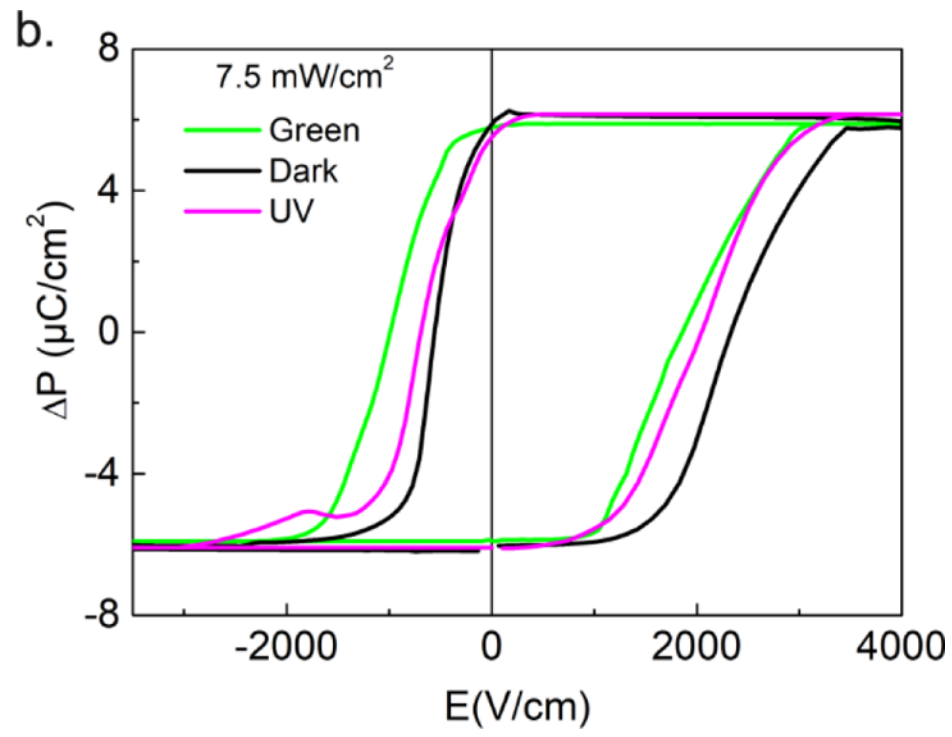

**Figure 5.** (a) Extracted nonswitching current for different light illumination conditions $C_s\frac{\partial V}{\partial t} \neq 0$; $\frac{\partial P_{ns}}{\partial t} \neq 0$, (b) polarization loop extracted for the condition $C_s\frac{\partial V}{\partial t} = \frac{\partial P_{ns}}{\partial t} = 0$.

The large electroresistive loop observed in the dark ($\Delta R \approx 290\%$) is strongly suppressed under UV light, decreasing to approximately 15% at a light intensity of 7.5 mW/cm$^2$ (Figure 3c). Continuous photogenerated charges enhance electron doping, shifting graphene's Fermi level ($E_F = \hbar v_F \sqrt{\pi n}$) toward the Dirac point, therefore increasing its resistance value at $R$(P2).[28,29] The nonswitching contribution with respect to the switching one becomes a less dominant source of resistive change, reducing the Fermi level variation from 0.134 to 0.012 eV (see Table S1 and details of the calculations in the Supporting Information).

In contrast, for the 530 nm green light ($E_g$ = 2.3 eV), the FE loop is much less affected for the same light intensity. Only shifts in coercive fields with a small effect on polarization are observed (Figure 3e).[30−32] This minimal effect on polarization modulation is also reflected in the graphene resistance (Figure 3f), where $R$(P2) remains less reduced with respect to $R$(P1), and the resistance difference ($\Delta R$) is reduced to only 229% under a light intensity of 7.5 mW/cm$^2$. The sample temperature measured with the thermal camera showed less than 0.2 K warming for the given light intensity. So, we can rule out the thermal effects to explain large and nonsymmetric effect on the $R$(V) loop of graphene. Instead, the observed graphene doping is governed by interfacial charge trapping and redistribution driven by light-induced domain dynamics. Even though the sample is saturated along [001], residual domains in other orientations may still exist and become mobile under illumination, resulting in charge redistribution at domain walls and interfaces. Consequently, even sub-band gap light can modify the local electrostatic potential "sensed" by graphene.[33,34]

The effect of illumination on the FE cycle is detailed in Figure 4 through the FE current, dynamic capacitance ($dQ/dV$), and the rate of resistance variation in graphene ($dR/dV$). In the dark, the current exhibits very low values and peaks at the coercive field, corresponding to polarization state switching, where atomic movement enhances the charge flow inside the FE. Under UV illumination, free charge generation increases, leading to a different photovoltaic current slope (Figure 4b), due to nonswitching contribution.[35] In contrast, green light does not significantly alter the current, indicating a lack of band-to-band electronic transitions of free charges generation (Figure 4f). Moreover, a reduction of the current loop opening with respect to dark loop indicates free charge reduction (trapping) under green light.

Dynamic capacitance provides insight into charge modulation and its distribution behavior within the FE material in response to the applied electric field. The capacitance, determined as the derivative of polarization charge with respect to the applied voltage ($dQ/dV$), exhibits peaks at the coercive electric field, where a significant accumulation of charge occurs at the FE surface.[36] Additionally, it takes on negative values when the electric field is reversed from its maximum, as shown in Figure 4c and its inset. A higher PV current requires a stronger electric field to counteract its effect and restore the positive capacitance in the device. In contrast, green light does not affect the negative capacitance region (Figure 4g) as green light does not generate a PV current (see Figure S3, in Supporting Information).

The variation of graphene resistance with gate voltage reflects charge transfer and the modulation of Fermi energy levels, which is influenced by the FE polarization. Under UV light illumination, the $dR/dV$ peak becomes smaller, confirming that enhanced charge screening reduces the effectiveness of ferroelectric polarization in modulating graphene resistance (Figure 4d) with peaks positions only slightly shifted. In contrast, for green light, the peak position is largely shifted while light-induced change in the peak magnitude is smaller (Figure 4h). Such behavior confirms different mechanisms of the photo response where switching and nonswitching contributions must be analyzed separately.

## 5. NONSWITCHING AND SWITCHING CONTRIBUTIONS FOR DARK, UV, AND GREEN LIGHT ILLUMINATIONS

In the case of light illumination, the analysis mentioned in Section 1 and Supporting Information, the nonswitching current has been extracted from as-measured FE loops as shown in Figure 5.

As can be seen, the UV light induces more free charges, leading to increased conductive response of the nonswitching part. On the contrary, the 530 nm light tends to reduce nonswitching part via charge trapping or thermal mechanisms (Figure 5a).[37] The switching only ferroelectric loops obtained after removing nonswitching part are depicted in Figure 5b. The loop under green light shifts toward negative electric fields making the loop more symmetric (abs ($E_c^+ + E_c^-$)) decreasing). At the same time, it reduces noticeably the remanent polarization. However, under UV the remanent polarization becomes even slightly larger and coercive fields shrink, indicating a different mechanism, under which becomes easier to switch the dipoles, as photogenerated carriers can screen depolarizing fields and unpin domain walls.

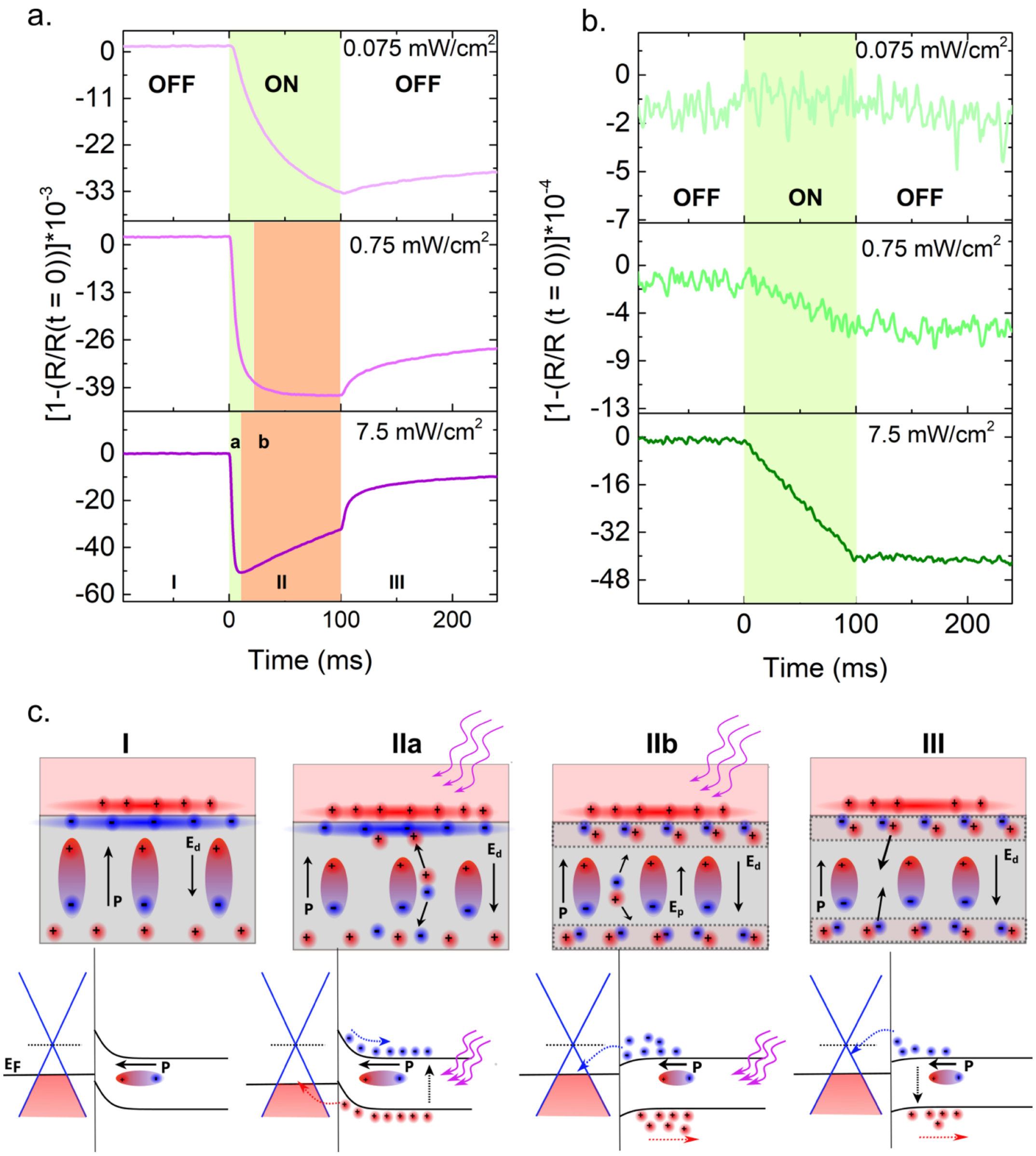

**Figure 6.** Graphene resistance variation for single (a) UV and (b) green light pulse of 100 ms duration each but of different intensities, and (c) schematic of charge dynamics and Fermi level modulation for different regions.

## 6. TIME EVOLUTION OF CHARGE DYNAMICS INSIDE THE FE AND ITS EFFECT ON THE GRAPHENE TRANSPORT

As follows from the analysis in the dark, the polarized state 1 (obtained by cycling the electric field from zero to 4 kV/cm and back to zero) consists of polarized dipoles and a minor contribution of free charges. To investigate the effect of light on a 100 ms time scale, the response of graphene resistance to UV and green light pulses is shown in Figure 6. The resistance was measured by applying a single 100 ms light pulse, while the FE substrate was in a positive remanent state P1. At low UV light intensity of 0.075 mW/cm$^2$ (Figure 6a), photogenerated free charges are driven by the internal field, modifying only the screening charges and causing a decrease in graphene resistance.[38] As light intensity increases to 0.75 mW/cm$^2$, a greater number of charge carriers are generated, neutralizing the screening charge cloud and forming a depletion layer, leading to resistance saturation.[39] Under the highest light intensity studied (7.5 mW/cm$^2$), the resistance initially decreases due to charge screening modulation (Figure 6a, lower panel; Region IIa) and then it starts to increase again while light is still ON (Figure 6a, lower panel; Region IIb). This behavior can be attributed to oppositely polarized photocarriers, which move against the internal electric field and generate a PV current, thereby leading to the observed increase in resistance in the IIb region (Figure 6a, lower panel). A schematic representation of charge movement within the ferroelectric and the resulting charge doping in graphene is illustrated in Figure 6c, highlighting the distinct response regions. Electronic bands become bent at the interface of graphene and FE depending on the polarization and photogenerated electrons and holes dope graphene according to the electric field in FE, as shown in Figure 6c for different regions mentioned above.

Green light exposure of the same 100 ms pulse does not generate free charges but instead most probably modifies electrostatics via photodomain effect, a process that occurs much more slowly compared to UV light case (Figure 6b). This gradual charge redistribution affects both screening and internal charges, exhibiting behavior similar to that observed in Figure 6a, Region IIa. Additionally, increasing green light intensity does not significantly alter the overall effect but slightly accelerates the process.

## 7. CONCLUSIONS

FE materials exhibiting a PV effect can demonstrate a wide range of photoresponse behaviors, depending on the illumination wavelength, intensity, and ferroelectric state. 2D materials, owing to their high sensitivity to nearby charge variations, are capable of amplifying and detecting even subtle changes in electrostatics. As a result, interfacing FEs with a 2D overlayer offers a promising platform for both sensing charge dynamics and inducing new functionalities in the 2D material via intrinsic doping modulation. We have reported that for low intensities, the beyond band gap (UV) illumination rapidly screens the FE polarization within a few milliseconds. For larger intensities, photogenerated carriers are able to organize into a current flow reducing the FE intrinsic depolarization field. In contrast, sub-band gap (green) light excitation induces a lower charge redistribution without significant free carrier generation within the FE material, leading to a gradual modulation of polarization, linear with light intensity. By separating switching from nonswitching pathways under controlled illumination, it is possible to directly quantify the competition between photovoltaic screening and other photo-induced properties which govern nonequilibrium ferroelectricity. These findings uncover light-controlled polarization screening in ferroelectric/2D heterostructures, enabling novel memory, neuromorphic, and logic devices. The described approach can be extended to more complex multicomponent interfaces, advancing understanding and design of functional heterostructures based on 2D materials.

**The Supporting Information:**

PE loop for leaky ferroelectric and after subtracting the leakage current; I−V characteristics of a ferroelectric with background linear capacitance and after subtraction; steady photovoltaic current observed for UV light and it is absent for green light; and parameters like Fermi energy shift in graphene, coercive field, screening charge, and charge density.

## AUTHOR INFORMATION

### Corresponding Authors

**Krishna Prasad Maity** − *Department of Physics, SRM University-AP, Amaravati 522240, India*; orcid.org/0000-0003-2307-1963; Email: kpmaity003@gmail.com

**Bohdan Kundys** − *Université de Strasbourg, CNRS, Institut de Physique et Chimie des Matériaux de Strasbourg, Strasbourg F-67000, France*; Email: kundys ipcms.fr

### Authors

**Mohd Uvais** − *Université de Strasbourg, CNRS, Institut de Physique et Chimie des Matériaux de Strasbourg, Strasbourg F-67000, France*

**Jean-François Dayen** − *Université de Strasbourg, CNRS, Institut de Physique et Chimie des Matériaux de Strasbourg, Strasbourg F-67000, France*

**Bernard Doudin** − *Université de Strasbourg, CNRS, Institut de Physique et Chimie des Matériaux de Strasbourg, Strasbourg F-67000, France*

**Roman Gumeniuk** − *Institut Für Experimentelle Physik, TU Bergakademie Freiberg, Freiberg 09596, Germany*; orcid.org/0000-0002-5003-620X

### Notes

The authors declare no competing financial interest.

## ACKNOWLEDGMENTS

We thank Fabien Chevrier and the staff of the STnano nanofabrication facility for daily support. K.M. thanks SRM University AP for supporting "seed grant" (SRMAP/URG/SEED/2024-25/054). We acknowledge the Agence Nationale de la Recherche for financial support through the grants MixDFerro (ANR-21-CE09-0029) and SOFIANE (ANR-23-CE09-0007) and IdEx Unistra (ANR 10 IDEX 0002), SFRI STRAT'US project (ANR 20 SFRI 0012), and EUR (QMat-ANR-18-EUR-0016) under the framework of the French Investments for the Future Program.

# Supporting Information

# Wavelength-Resolved Control of Photovoltaic Screening and Defect-Mediated Doping in Photo-Ferroelectric/Graphene Devices

Krishna Prasad Maity[1]*, Mohd Uvais[2], Jean-François Dayen[2], Bernard Doudin[2], Roman Gumeniuk[3], and Bohdan Kundys[2]*

[1]*Department of Physics, SRM University, Amaravati-522240 Andhra Pradesh, India*

[2]*Université de Strasbourg, CNRS, Institut de Physique et Chimie des Matériaux de Strasbourg, UMR 7504, Strasbourg F-67000, France*

[3]*Institut für Experimentelle Physik, TU Bergakademie Freiberg, Freiberg 09596, Germany*

Email: kpmaity003(AT)gmail.com

Email: kundys(AT)ipcms.fr

**Compensation for non-switching currents:**

**1. Leakage current compensation.**

The insulating imperfection of many dielectric materials leads to electric leakage, allowing DC current to flow, slowly discharging the capacitor. Such semiconducting behavior of a bad insulator naturally reflects itself in the increment of the current value as a function of the applied electric field, while no current flow is expected for a good insulator[1] (fig. S1a). The integration of the curves with respect to time is nothing like the ferroelectric hysteresis[2,3]. In ferroelectric materials, however, an additional switching contribution to the current (a peak) is expected to arise from electric dipole movement (eq.1, in main text, absence of stray capacitance, $C_s \frac{\partial V}{\partial t} = 0$). This ferroelectric contribution to the often non-linear leakage current gives a typical lossy uncompensated ferroelectric (FE) loop, as shown in Fig. S1b (blue curve, before leakage subtraction).

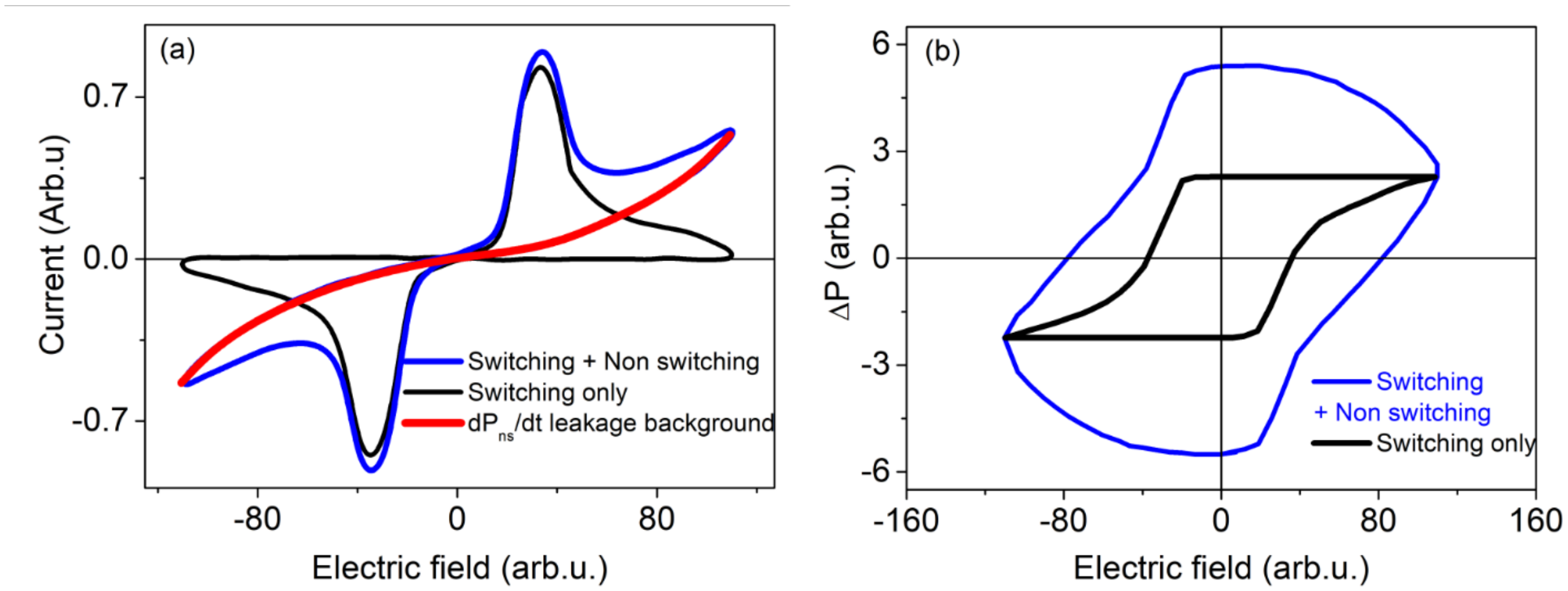

Fig. S1: Example of the I-E characteristic recorded for electrically leaky ferroelectric material (a) and resulted P-E loops before and after the leakage subtraction (b).

It has to be noted that I-V characteristic represents the model of a very first loop taken for non-single domain ferroelectric sample. Indeed, if the electric field is applied stepwise in a parallel direction of the forced polarization (sample is already polarized) no switching ferroelectric current occurs in this direction. Consequently, additional cycle has to be taken, to present the two consecutive peaks in the positive and negative voltages. This effect is central in the “double wave method” to determine nonlinear the leakage current background (bold red curve in the fig. S1a). As one can see, the electric leakage manifestly increases both the apparent coercive fields and apparent remanent polarization. In order to avoid the leakage current contribution additional DC conductivity compensating circuit can be added to the electric circuit. Alternatively, and more easily one can propose simple subtraction of background leakage current from the I-V curve as shown on figure S1a. The application of this procedure gives a clear saturation of the ferroelectric loop that is also shown on figure S1(b) (black curve) for comparison.

## 2. Stray capacitance compensation.

To measure the ferroelectric loop, the sample is placed into an electrical circuit of the DC electrometer together with its electrical wires. If the capacitance of the sample is smaller than or equal to the stray capacitance of the electrical circuit, the stray capacitance should be taken into account. This extra capacitance usually has detrimental effects on the operation of "real-life" circuits, changing the shape of ferroelectric loops, therefore also requiring compensation. Additionally, due to stray capacitance ($C_s\frac{\partial V}{\partial t} \neq 0$; $\frac{\partial P_{ns}}{\partial t} = 0$ in main text eq.2), the dipole reversibility time may vary, consequently affecting both light-induced and dark polarization studies. The presence of stray capacitance reflects itself in the shape characteristic of both I-V and ferroelectric loops, as shown in Figure S2(a) opening a gap in I(V) measurements.

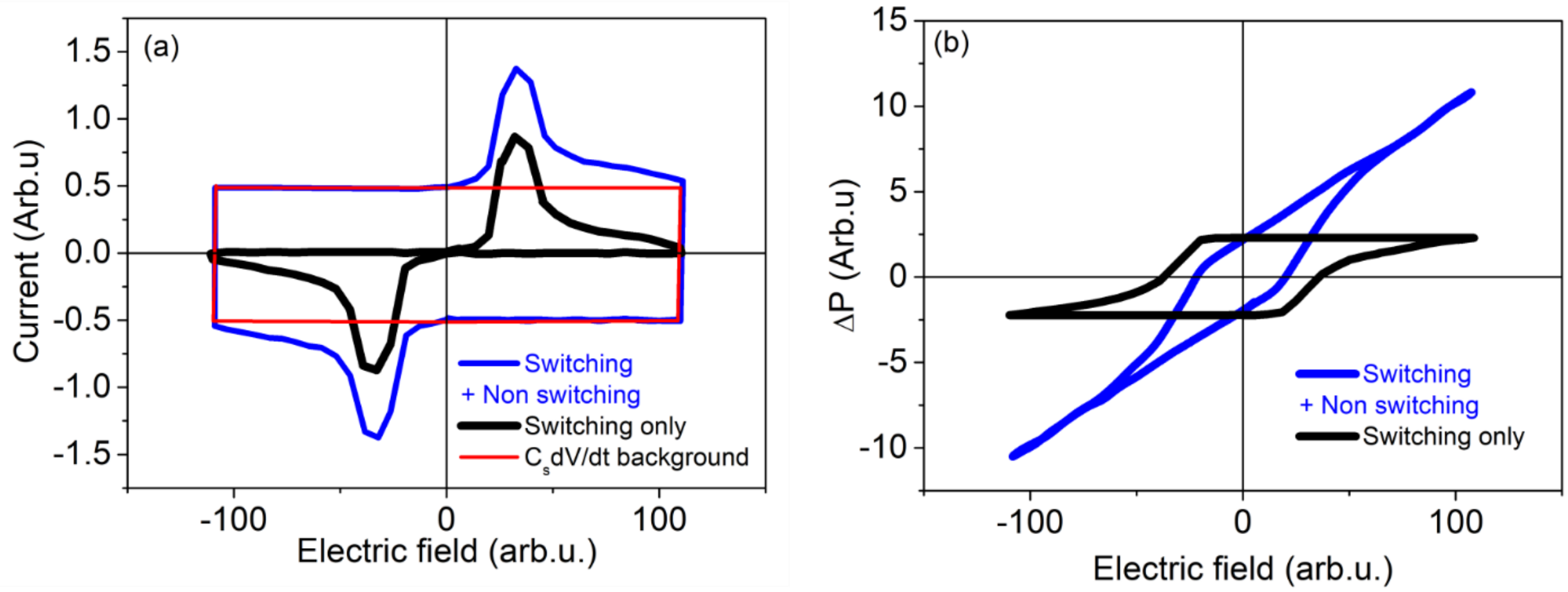

Fig. S2: Example of the I-V characteristic recorded on a (a) background large linear capacitance contribution and (b) same I-V curve after subtracting the linear capacitance compensation.

As one can see from Figure S2 (b), the stray capacitance largely decreased the apparent coercive fields and can be successfully removed with this straightforward approach.

The presence of both aforementioned contributions (see eqs. 2 and 3 in the main text), often the occur in real samples, and they can be removed using this approach. The usage of a quasi DC ferroelectric test unit offers advantage over conventional AC ferroelectric measurements due to low electric fatigue risk due to ultra-low frequency tracing. This is of great importance for photoelectric investigations. A more accurate, simple, theoretical compensation for stray linear capacitances and dielectric leakage current can not only be used for correct polarization and coercivity estimation, but also for investigation of intrinsic origins of non-switching currents in functional FE/2D heterostructures.

### 3. PV Effect for UV and Green light Illumination

PV current measurements reveal the non-zero current under UV, absent for 530 nm green light (Fig. S3). Above band gap optical excitation creates free charge carriers to flow inside ferroelectric and generate bulk-photovoltaic current, which is not present for the sub-bandgap excitations.

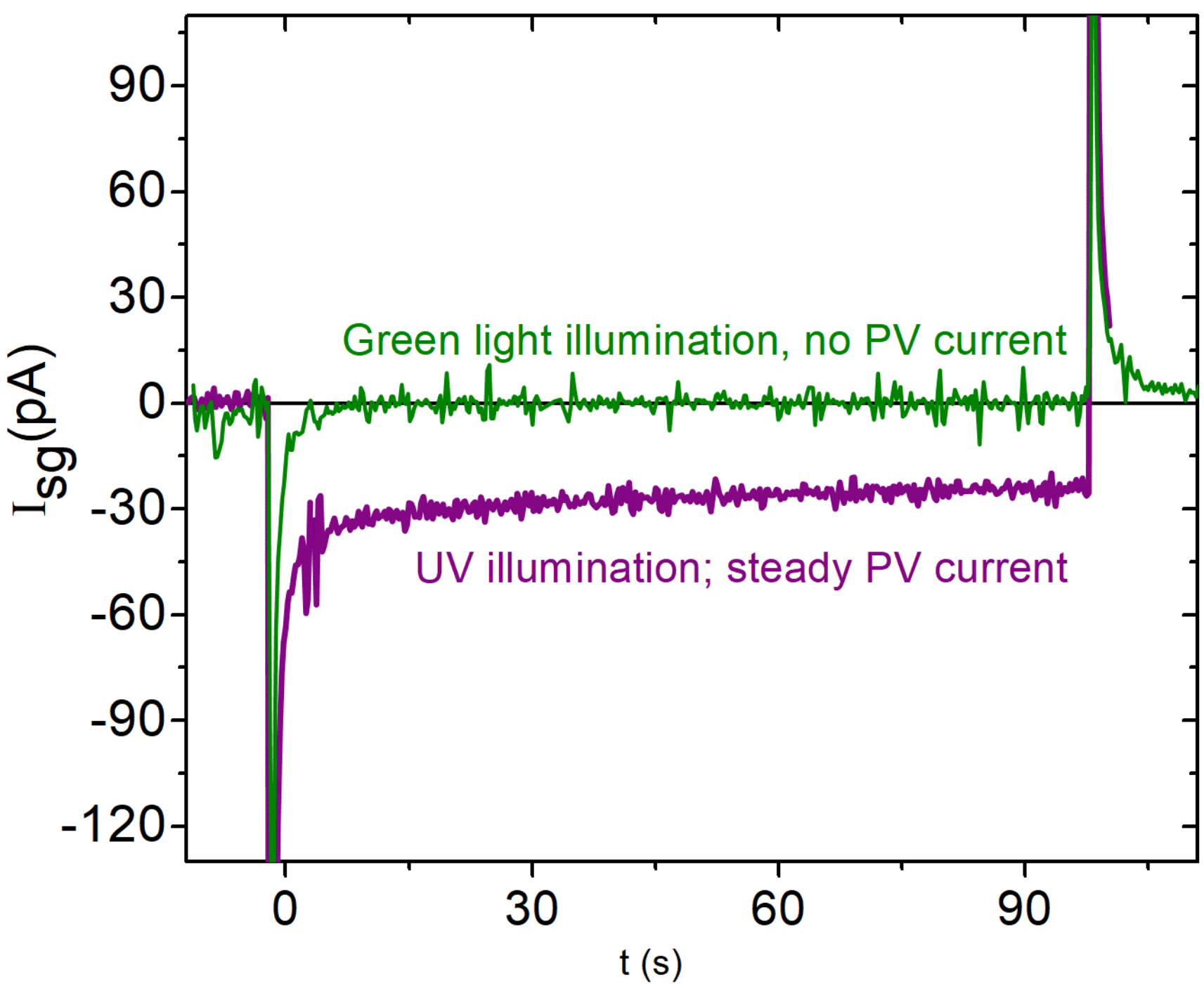

Fig. S3: The PV current is shown under UV light illumination, while its steady state value is negligible under green light illumination.

**4. Calculations for table S1:**

The FE built-in field is given as:

$E_{int} = (E_c^+ + E_c^-)/2$; where $E_{int}$ is the depolarization field, $E_c^+$ and $E_c^-$ are the coercive fields.

$E_{int} = -(P-\sigma_s)/\varepsilon_0\varepsilon_r$; P is saturation polarization, $\sigma_s$ is the screening surface charge density inside the ferroelectric, $\varepsilon_0 = 8.85\times10^{-12}$ F/m is free space permittivity and $\varepsilon_r$ is the relative permittivity

So, screening charge density inside ferroelectric can be written as

$\sigma_s = P + \varepsilon_0\varepsilon_r E_{int}$

The relative permittivity of PMN-PT is calculated from the dQ/dV graph where its value remains constant near E = 0 V/cm. Using the equation $\varepsilon_r = (dQ/dV)*(d/A.\varepsilon_0)$; d is the width, and A is the surface area of ferroelectric.

The effective gate voltage becomes enhanced by the polarization of ferroelectric and effective gate follows the equations

$V_{eff} = V_G + Pd/\varepsilon_0\varepsilon_r A$; $n = (\varepsilon_0\varepsilon_r A/ed)\, V_{eff}$, where n is the charge density of graphene and e = $1.6\times10^{-19}$ Coulomb.

Charge density defines the Fermi levels of graphene by the equation

$E_F = \hbar v_F \sqrt{\pi n}$

Where $E_F$ is the Fermi energy, $v_F$ is the Fermi velocity of the order of $10^6$ m/s.

Mobility of the charge can be calculated from the constant region of ($dI_{ds}/dV_G$ ) graph

$\mu = (L/W)*(1/C)*(1/V_{ds})*(dI_{ds}/dV_G)$

μ is the mobility of charge carriers, L is length, W is width and $V_{ds}$ is the voltage across source and drain.

***For our sample,***

L = 0.5 cm, W = 0.05 cm and d = 0.3 mm

In dark (Light intensity = 0 mW/cm$^2$)

$E_c^+$ = 2373 V/cm; $E_c^-$ = -638 V/cm; P1 = 7.5 μC/cm$^2$ and P2 = -7.2 μC/cm$^2$

So, $E_{int}$ = 867 V/cm, using $\sigma_s$ = 11.2 μC/cm$^2$ (considering as polarization at P1)

C = $3.53\times10^{-9}$ Farad and $\varepsilon_0 = 8.85\times10^{-12}$ F/m, Using $\varepsilon_r = 4.79\times10^4$

The charge density of graphene is calculated at P1

n (P1) = (L/W)*(1/R(P1)eμ) = $5.36\times10^{12}$ /cm$^2$, Using $E_F$ = 0.134 eV and considering $v_F = 10^6$ m/s

The value of $(1/V_{ds})*(dI_{ds}/dV_G)$ is calculated from the constant region around $V_G$ = 0 V.

The mobility can be estimated as, μ = 6860 cm$^2$V$^{-1}$s$^{-1}$

Similarly, parameters for UV light illumination are calculated for the below table.

**Table S1:** The variation of coercive field, remanent polarization, screening charge density, resistance of graphene at two remanent states and Fermi energy differences for UV light incident.

| Light Intensity (mW/cm$^2$) | $E_c^+$ (V/cm) | $E_c^-$ (V/cm) | $E_{int}$ (V/cm) | P1 (μC/cm$^2$) | P2 (μC/cm$^2$) | $\sigma_s$ (μC/cm$^2$) | R (P2) (kΩ) | R (P1) (kΩ) | n (P2) $\times 10^{12}$ (cm$^{-2}$) | n (P1) $\times 10^{12}$ (cm$^{-2}$) | $E_F$ (P2) (meV) | $E_F$ (P1) (meV) | $\Delta E_F$ (meV) |
|---|---|---|---|---|---|---|---|---|---|---|---|---|---|
| 0 | 2373 | - 638 | 867 | 7.5 | - 7.2 | 11.2 | 1.7 | 6.7 | 5.36 | 1.36 | 269 | 136 | 134 |
| 7.5 | 2237 | - 1109 | 564 | 12.8 | - 9.6 | 15.2 | 4.7 | 4.6 | 2.28 | 1.98 | 176 | 164 | 12 |